\documentclass[aps,prl,reprint,showpacs,groupedaddress]{revtex4-1}

\usepackage{amsmath}
\usepackage{amssymb}
\usepackage{graphicx}
\usepackage[capitalize]{cleveref}
\usepackage{color}

\def\pl{\partial}

\def\al{\alpha}
\def\bt{\beta}
\def\Ga{\Gamma}
\def\ga{\gamma}
\def\de{\delta}

\def\te{\theta}

\def\lam{\lambda}

\def\om{\omega}
\def\ep{\epsilon}

\def\l{\left (}
\def\r{\right )}

\def\fr{\frac}
\def\la{\label}

\def\vs{\vspace}

\def\ran{\rangle}
\def\lan{\langle}

\def\tl{\tilde}
\def\tm{\times}

\begin{document}

\newcommand{\ba}[1]{\begin{array}{#1}} \newcommand{\ea}{\end{array}}



\def\Journal#1#2#3#4{{#1} {\bf #2}, #3 (#4)}

\def\NCA{\em Nuovo Cimento}
\def\NIM{\em Nucl. Instrum. Methods}
\def\NIMA{{\em Nucl. Instrum. Methods} A}
\def\NPB{{\em Nucl. Phys.} B}
\def\PRL{\em Phys. Rev. Lett.}
\def\PRD{{\em Phys. Rev.} D}
\def\ZPC{{\em Z. Phys.} C}

\def\st{\scriptstyle}
\def\sst{\scriptscriptstyle}
\def\mco{\multicolumn}
\def\epp{\epsilon^{\prime}}
\def\vep{\varepsilon}
\def\ra{\rightarrow}
\def\ppg{\pi^+\pi^-\gamma}
\def\vp{{\bf p}}
\def\ko{K^0}
\def\kb{\bar{K^0}}
\def\al{\alpha}
\def\ab{\bar{\alpha}}

\def\np{Nucl. Phys. {\bf B}}
\def\mpl{Mod. Phys. {\bf A}}\def\ijmp{Int. J. Mod. Phys. {\bf A}}
\def\cmp{Comm. Math. Phys.}\def\prd{Phys. Rev. {\bf D}}

\def\oa{\bigcirc\!\!\!\! a}
\def\ob{\bigcirc\!\!\!\! b}
\def\oc{\bigcirc\!\!\!\! c}
\def\oi{\bigcirc\!\!\!\! i}
\def\oj{\bigcirc\!\!\!\! j}
\def\ok{\bigcirc\!\!\!\! k}
\def\ve{\vec e}\def\vk{\vec k}\def\vn{\vec n}\def\vp{\vec p}
\def\vv{\vec v}\def\vx{\vec x}\def\vy{\vec y}\def\vz{\vec z}

\newcommand{\AdS}{\mathrm{AdS}}
\newcommand{\dd}{\mathrm{d}}
\newcommand{\eee}{\mathrm{e}}
\newcommand{\sgn}{\mathop{\mathrm{sgn}}}

\def\a{\alpha}
\def\b{\beta}
\def\g{\gamma}

\newcommand\lsim{\mathrel{\rlap{\lower4pt\hbox{\hskip1pt$\sim$}}
    \raise1pt\hbox{$<$}}}
\newcommand\gsim{\mathrel{\rlap{\lower4pt\hbox{\hskip1pt$\sim$}}
    \raise1pt\hbox{$>$}}}

\newcommand{\beq}{\begin{equation}}
\newcommand{\eeq}{\end{equation}}
\newcommand{\bea}{\begin{eqnarray}}
\newcommand{\eea}{\end{eqnarray}}
\newcommand{\noi}{\noindent}


\title{Higgs-Squark-Slepton Inflation from the MSSM}

\author{Zurab Tavartkiladze}
\email[]{zurab.tavartkiladze@gmail.com}

\affiliation{Center for Elementary Particle Physics, ITP, Ilia State University, 0162 Tbilisi, Georgia}


\begin{abstract}

The inflation within the MSSM is proposed, where the inflaton field is a combination of the Higgs, squark and
slepton states.  While the inflationary phase is fully governed by the electron Yukawa superpotential coupling,
the fields' condensates float along the flat $D$-term trajectory. This predicts the
MSSM parameter $\tan \beta \simeq 13.1$ determined via the value of the curvature perturbation amplitude.
The values of the scalar spectral index and the tensor-to-scalar ratio
are predicted to be $n_s\simeq 0.966$ and $r=0.00118$.
The postinflation reheating of the Universe proceeds by the radiative decay of the inflaton to the two gluons
($\phi \to gg$) with the reheating temperature  $T_r\simeq 1.4\cdot 10^7$~GeV.

\end{abstract}



\maketitle

\section{I. Introduction}
\vs{-0.3cm}
Besides the firm support from the Planck Collaboration measurements \cite{Akrami:2018odb},  the inflationary paradigm \cite{first-infl} has strong theoretical motivations.
It elegantly solves many problems of the big bang cosmology.
It is very motivated and also, as it turns out, highly challenging to build successful inflation which has
close connection to the particle physics model whith consistent construction.
For this purpose, the supersymmetric (SUSY) setup (insuring flatness of the inflaton's potentiol protected by the supersymmetry)
looks one of the best choice \cite{susy-infl}  and
the Standard Model's minimal SUSY extension  (MSSM) seems the reasonable framework to deal with.
However, most of the successful inflation models exploit additional MSSM singlet(s).
 Among those  are models of inflation within NMSSM framework \cite{NMSSM-infl}, which  although
motivated by theoretical and phenomenological reasons, have reduced predictive power because of additional parameters.
Inflation models with MSSM field content exploiting slepton and/or squark states along $D$-term flat directions has been
studied in numerous works \cite{Allahverdi:2006iq} but with utilizing higher order operators involving new free parameters in the inflation process.
Note that the successful inflation within various well motivated extensions of the MSSM, such as SUSY GUTs and SUSY left-right symmetric models,
have been considered \cite{GUT-infl}. However, still, all these constructions involve additional MSSM singlet states with additional couplings.

 In a recent paper \cite{Tavartkiladze:2019cfb} within the MSSM,  the model of inflation along $D$-flat trajectory was proposed,
 where inflaton field emerged as a combination of the slepton and Higgs fields.
 The model utilized nonminimal  K\" ahler potential, however,
 in the inflation and postinflation reheating processes only MSSM Yukawa
   superpotential couplings have been involved. This made the model very predictive. In this paper we pursue this approach and
   investigate the possibility of involvement of the squark (the superpartners of the quarks) states into the inflation process.
   We present an interesting and novel possibility in which inflaton emerges as a superposition of the Higgs, squark and slepton 
   states. The inflaton potential, emerged from the superpotential $F$-term, involves only the electron Yukawa coupling.
   This fixes the value of the MSSM parameter $\tan \beta $. Besides this, the inflaton decay and subsequent reheating process
   is fully governed by the known MSSM couplings. Thus, very close interconnection between cosmology and particle physics model
   is established.

   Successful inflation is realized by the specific form of nonminimal  K\" ahler  potential. Proposed inflation model,
   which is disucussed and investigated in next two sections,
   also has several interesting phenomenological implications (discussed at the end of the paper).

\section{II. The framework and the inflaton potential}
\vs{-0.3cm}

The framework we are using is the $N=1$  supergravity \cite{sugra-pot,Kugo:1982mr}.
The action is built up from the $D$ and $F$-term Lagrangian densities
\begin{eqnarray}
{\cal L}_{D}+{\cal L}_F~,
\la{sugraL}
\end{eqnarray}
which  are determined by the K\" ahler potential ${\cal K}$, the superpotential $W$ and
 by the gauge kinetic function $f_{IJ}$. By the superconformal formulation, the ${\cal L}_{D, F}$ are given as follows  \cite{Kugo:1982mr}:
\begin{eqnarray}
&{\cal L}_{D}=- 3\int \!\!d^4\theta \,e^{-{\cal K}/3}\bar{\phi}^\dag \bar{\phi }, \nonumber \\
&{\cal L}_F=\int \!\!d^2\theta\bar{\phi }^3\, W +\frac{1}{4}\int \!\!d^2\te f_{IJ}{{\cal W}}^{\alpha I}{{\cal W}}_{\alpha}^J+ {\rm h.c.}
\label{DFlagr}
\end{eqnarray}
where  $\bar{\phi }$ is the conformal compensator chiral superfield. The ${{\cal W}}^{\alpha I}$ denote the gauge chiral superfield corresponding to the $SU(3)_c$, $SU(2)_L$ and $U(1)_Y$ symmetries. We will consider $f_{IJ}=\de_{IJ}$ - the canonical kinetic terms  for the gauge superfields.
In (\ref{DFlagr}) and below, where it is convenient, we set  the reduced Planck mass $M_{Pl}$ to one.
 In this way, any dimensionful quantity will be understood to be measured in the unit(s) of  $M_{Pl}$($=2.4\tm 10^{18}$~GeV).

The  K\" ahler potential  ${\cal K}$ and the superpotential  $W$ are the functions of the  MSSM  chiral superfields $\Phi_I$.
The latter are three families of quark and lepton superfields $\l q, u^c, d^c, l, e^c\r_{\al }$ ($\al =1,2,3$ is the family index),
and the up and down-type Higgs doublet chiral superfields $h_u, h_d$:
\begin{eqnarray}
\Phi_I=\left \{ \l q, u^c, d^c, l, e^c\r_{\al }, h_u, h_d \right \} .
\la{states}
\end{eqnarray}
After integrating out the auxiliary fields and fixing the conformal symmetries $\bar{\phi}=1$, the scalar potential
will get contributions from $F$ and $D$-terms:
\begin{eqnarray}
V=V_F+V_D,
\la{VFD}
\end{eqnarray}
where the $F$-term scalar potential is given by  \cite{sugra-pot,Kugo:1982mr}:
\begin{eqnarray}
& V_F=\,e^{{\cal K}}\left(D_{\bar J}{\bar W}{\cal K}^{{\bar J}I}D_IW - 3|W|^2\right),
\label{sugrVF}
\end{eqnarray}
where $D_IW\!=\!(\fr{\pl }{\pl \Phi_I}+\fr{\pl {\cal K}}{\pl \Phi_I})W$
and $D_{\bar J}{\bar W}\!=\!( \fr{\pl }{\pl \Phi_J^\dag }+\fr{\pl {\cal K}}{\pl \Phi_J^\dag }){\bar W}$.
The matrix ${\cal K}^{{\bar J}I}$ is an inverse of the   K\" ahler `metric'  ${\cal K}_{I{\bar J}}=\fr{\pl^2 {\cal K}}{\pl \Phi_I \pl \Phi_I^\dag }$. Thus,
 ${\cal K}_{I{\bar M}}{\cal K}^{{\bar M}J}=\de_I^J$ and ${\cal K}^{{\bar I}M}{\cal K}_{M{\bar J}}=\de^{\bar I}_{\bar J}$.

Further, we will use the following  nonminimal K\" ahler potential:
\begin{eqnarray}
\la{totalK}
&{\cal K}=-\ln (1-\sum_{I}\Phi_I^\dag e^{-V}\Phi_I) ~,
\end{eqnarray}
which in the small field limit ($\Phi_I\!\ll 1$) has the canonical form ${\cal K}\to \sum_{I}\Phi_I^\dag e^{-V}\Phi_I$.
However, for the large values of the fields,  as was shown \cite{Tavartkiladze:2019cfb}, the form of
(\ref{totalK}) together with the MSSM Yukawa superpotential terms can give successful inflation
 with observables determined in terms of the MSSM parameters.
Note that  with logarithmic but slightly different
 K\" ahler potential (exploiting MSSM singlet states), the successful chaotic inflation was  realized
 in  Refs. \cite{Kallosh:2013hoa, Ferrara:2016fwe}.
 In this paper, we study the inflation with the inflaton emerging from the scalar components of the MSSM states only.
We will be focusing to realize inflation along the flat $D$-term trajectory, i.e. $\lan V_D\ran =0$ during the inflation.
In works \cite{Affleck:1984fy,Dine:1995kz}
the  slepton and/or squark condensates along the flat directions have been used for the baryogenesis process in the early Universe.
The inflation with  sleptons and/or squarks has been studied in Refs. \cite{Allahverdi:2006iq},
 however, these constructions  exploit higher order operators with many new parameters involved in the inflation process.

 With the K\" ahler potential (\ref{totalK}), the $D$-term potential $V_D$ is build from the Killing potentials ${\cal D}_G$
 corresponding to the $U(1)_Y, SU(2)_L$ and $SU(3)_c$ gauge symmetries [$G=Y, SU(2), SU(3)$]:
 \beq
V_D=\fr{g_1^2}{8}{\cal D}_Y^2+\fr{g_2^2}{2}({\cal D}^i_{SU(2)})^2+\fr{g_3^2}{2}({\cal D}^a_{SU(3)})^2.
\la{VD-pot}
\eeq
The Killing potentials ${\cal D}_G$ are related to the $D$-terms as
\beq
{\cal D}_G=\fr{D_G}{1-\sum_{I}\Phi_I^\dag \Phi_I}
\la{kilingD-D}
\eeq
where $\Phi_I$ in (\ref{kilingD-D}) stand for lowest scalar component of the corresponding chiral superfield.
On the other hand, the $D$-terms $D_G$ corresponding to the $U(1)_Y$, $SU(2)_L$ and $SU(3)_c$, are respectively:
\begin{eqnarray}
\label{D-terms}
&D_Y=|h_d|^2-|h_u|^2-2|\tl e^c_{\al}|^2+|\tl l_{\al}|^2
\nonumber \\
&-\fr{1}{3}|\tl q_{\al}|^2+\fr{4}{3}|\tl u^c_{\al}|^2-
\fr{2}{3}|\tl d^c_{\al}|^2 ~,
\nonumber \\
&D^i_{SU(2)}=\!\fr{1}{2}\!\l h_d^\dag \tau^ih_d-h_u^\dag \tau^ih_u+\tl{l}_{\al }^{\dag } \tau^i\tl l_{\al}
+ \tl{q}_{\al }^{\dag } \tau^i\tl q_{\al}\r ,
\nonumber \\
&D^a_{SU(3)}=\!\fr{1}{2}\!\l \tl{q}_{\al }^{\dag } \lam^a\tl q_{\al}-\tl{u}_{\al }^{c\dag } \lam^a\tl u^c_{\al}-
\tl{d}_{\al }^{c\dag } \lam^a\tl d^c_{\al} q_{\al}\r .
\end{eqnarray}
 In (\ref{D-terms}) the summation under the family index  $\al=1,2,3$  is assumed. $\tau^i/2$ and $\lam^a/2$ are respectively
$SU(2)_L$ and $SU(3)_c$ generators  ($i=1,2,3$, $a=1,\cdots ,8$).

One can readely check that given by (\ref{kilingD-D}), (\ref{D-terms})
the equations
\begin{eqnarray}
&{\cal K}_{I\bar J}(iT^A)^I_M\Phi_M=i\fr{\pl }{\pl \Phi_J^\dag }{\cal D}^A, \nonumber \\
&{\cal K}_{I\bar J}(iT^A)_I^M\Phi_M^\dag=i\fr{\pl }{\pl \Phi_I}{\cal D}^A
\la{Kiling-eqs}
\end{eqnarray}
are automatically satisfied ($T^A$ stand for the generator/charge of the corresponding gauge symmetry).
As known from supergravity constructions \cite{sugra-gauge},
these ensure the consistent supergravity gauge invariance.
\vs{-0.3cm}
\subsection{II.1.  Choice of  Flat $D$-term Direction}
\vs{-0.3cm}

In MSSM there are numerous solutions with the
$D$-term  flat directions, which have been classified in \cite{Gherghetta:1995dv}.
Here we consider one (the $e^clqu^c$-type flat direction) involving the scalar component of $h_d$,
the sleptons $\tl e^c, \tl l$ and the squarks $\tl q, \tl u^c$.
The state $h_d$, different families of sleptons, squarks (of the quantum numbers indicated above)
will share vacuum expectation values (VEVs) by appropriate weights.
 In particular, we will consider the following VEV configuration:
\vs{-0.6cm}
\begin{eqnarray}
&\lan \tl e^c_1\ran \!=\!z,~~~
\! \lan h_d\ran \!=\! \left(\begin{array}{c}
 zc_{\te }
\\
0
\end{array}\right),~~~
\! \lan \tl l_2\ran \!=\! \left(\begin{array}{c}
zs_{\te }
\\
0
\end{array}\right)\!\!
\begin{array}{c}
 \\ \!\!{~}_{\uparrow }\!\! \\ \vspace{-0.1cm} {~}_{\!SU(2)_L}\!\! \\ \vspace{0.6cm}{~}_{\downarrow } ~
 \end{array}~ \nonumber  \\
&\begin{array}{ccc}
 & {\begin{array}{ccc}
 &~~  ~~  {~}_{\leftarrow ~SU(3)_c ~\rightarrow }&\\
\end{array}}\\
&{\! \lan \tl q_1\ran = \left(\begin{array}{ccc}

 0& ~~ 0  & ~~ 0
\\
  0 &~~ 0 & ~~z
\end{array}\right)}
\end{array}  \!\!\!\!
\begin{array}{c}
 \\ {~}_{\uparrow } \\ \!  {~}_{SU(2)_L}\! \\ \vspace{0.3cm}{~}_{\downarrow } ~
 \end{array}~\nonumber \\
&\lan \tl u^c\ran =\! \l \begin{array}{ccc}
  0, &~ 0, & ~zc_{\varphi }
\end{array} \!\r ,~~~~~\nonumber \\
&\lan \tl t^c\ran = \!\l \begin{array}{ccc}
  \!0, & 0, & zs_{\varphi }e^{i\om }\!
\end{array} \!\r ,~~~~~
\label{VEVs}
\end{eqnarray}
where actions of $SU(3)_c$ and $SU(2)_L$ are depicted  schematically.
Also, the short handed definitions $\cos (\te, \varphi ) \equiv c_{\te, \varphi}$ and $\sin(\te, \varphi )\equiv s_{\te, \varphi}$ are
used. The angles $\te , \varphi $ and the phase $\om $ will be determined/fixed from the superpotential.
Essential point is the fact that
with (\ref{VEVs}) configuration (and with zero VEVs of all remaining fields),
all $D$-terms [of Eq. (\ref{D-terms})] vanish (and thus $\lan V_D\ran =0$)
 for arbitrary values of $z$, $\te, \varphi $ and $\om $. While the values of $\te, \varphi , \om $
 will be fixed, the $z$ will be a dynamical variable and will be related to the inflaton field.
As will be shown, this  will lead to the predictive and successful inflation.
To see how things work out, we need to consider the superpotential couplings.

\vs{-0.3cm}
\subsection{II.2.  The Superpotential and Inflaton Potential}
\vs{-0.3cm}
The MSSM superpotential includes three $Y_E, Y_D$ and $Y_U$ Yukawa matrix-couplings  and the $\mu $-term:
\begin{eqnarray}
\label{WMSSM}
&W_{\rm MSSM}\!=\!e^cY_Elh_d+ qY_Dd^ch_d+ qY_Uu^ch_u+\mu h_uh_d .~~~~
\end{eqnarray}
Without loss of any generality, we choose the field basis such that
the Yukawa matrices are:
\begin{eqnarray}
\label{Yuk-basis}
&Y_E=Y_E^{\rm Diag}={\rm Diag}\l \lam_e, \lam_{\mu }, \lam_{\tau }\r ,
\nonumber \\
&Y_D=Y_D^{\rm Diag} ,~Y_U=V_{CKM}^TY_U^{\rm Diag}.
\end{eqnarray}
From Eq. (\ref{WMSSM}), with (\ref{VEVs})  we have:
\beq
F_{e^{-}}^* =-\lam_ez^2c_{\te } ~.
\la{Fterms}
\eeq
In our construction, this will be the only nonvanishing $F$-term contributing to the inflation potential.
As mentioned, the $\te, \varphi $ and $\om $ will be fixed from the superpotential couplings by imposing the vanishing conditions
for all remaining $F$-terms.
For instance, the requirement $F_{h_u^{(2)}}=0$ gives the condition
$z^2\l V_{ud}\lam_uc_{\varphi }+ V_{td}e^{i\om }\lam_ts_{\varphi }\r =0$, which is satisfied by fixing $\om $ and $\varphi $ as follows:
$\om =\pi+{\rm Arg }\l \fr{V_{ud}}{V_{td}}\r $, $\tan \varphi \!=\!\fr{\lam_u}{\lam_t}\left |\fr{V_{ud}}{V_{td}} \right |\simeq 3\cdot 10^{-4}$ \cite{phi-val}.
Here and below,  the $\mu $-term being too small($\sim $few TeV) and therefore irrelevant for inflation, will be ignored.
Moreover, we include additional superpotential $W'$,  which ensures fulfilment of the condition $F_{d^c}=0$.
Two cases - {\bf (i)} and {\bf (ii)} - can be considered which have different low energy implications, but lead to the
 same inflation process.

{\bf (i)} $W'=-\lam q_1l_2d^c$.
This coupling, together with the couplings (\ref{WMSSM}) gives $\lan F_{d^c}^*\ran =z^2(-\lam_dc_{\te}+\lam s_{\te})=0$, i.e. fixing the angle $\te $
as $\tan \te =\fr{\lam_d}{\lam }$.

{\bf (ii)} $W'=\lam e^c_1(q_1l_2u^c)(q_1h_dd^c)$ which gives $\lan F_{d^c}^*\ran =z^2c_{\te}(-\lam_d+\lam z^4c_{\varphi }s_{\te})=0$ and fixes the angle $\te $ as follows $s_{\te}\simeq \fr{\lam_d}{\lam z^4}$.

 For both these cases we will be considering  the  suppressed values of $\te <0.1$ (i.e. $c_{\te}\simeq 1$), therefore expressions given above are pretty accurate \cite{te-approx}.

It is essential and very important that the values of $\te , \varphi $ and $\om $ are fixed. Since they are parameterizing the field configuraation along the $D$-term flat direction [see Eq. (\ref{VEVs})] they are dynamical degrees and their fixation means their stabilization. This ensures that during the inflation there are no unstable/runaway directions. Plots in Fig.\ref{plot-1}
 represent the potential as a function of $(\phi,~ \te )$ and $(\phi,~ \varphi )$ variables respectively [$\phi $
  denotes inflaton and is related to $z$ via Eq. (\ref{z-t}). See also the caption of Fig.\ref{plot-1}].
They demonstrate that the valley (with a slight slope) is along the direction of the inflaton field $\phi$. Also, it is important that there are no other tachyonic or fast moving
degrees of freedom. We have checked, with the couplings and arrangements given above, and made sure that the presented inflation scenario is fully consistent.

Now we are ready to  derive the inflaton potential.
With the VEV configuration (\ref{VEVs}) we have $\lan W\ran =0$ and nonvanishing
$F$-term of Eq. (\ref{Fterms})  gives from (\ref{sugrVF}):
\begin{eqnarray}
&V_F=e^{\cal K} {\cal K}^{{e^{-}}^\dag e^{-}}|F_{e^{-}}|^2,
\la{V}
\end{eqnarray}
which depend on the form of the ${\cal K}$.
Had we have considered minimal (canonical) form for the  K\" ahler potential $\sum_{I}\Phi_I^\dag e^{-V}\Phi_I$,
with (\ref{Fterms}) and $\te \ll 1$, the inflaton potential would be $\lam_e^2z^4$. The latter would give an unacceptably large
 tensor-to-scalar ratio. Thus, refuting this possibility, we are considering the form given by
 Eq. (\ref{totalK}).
 The kinetic part, which includes $(\pl z)^2$ is
 \begin{eqnarray}
&{\cal K}_{I\bar J}\pl \Phi_I \pl \Phi_J^{*}\to (\pl V_z)^\dag \lan {\cal K}(z) \ran \pl V_z,
\la{kin-z}
\end{eqnarray}
where with (\ref{totalK})  and  (\ref{VEVs}) we have:
\begin{eqnarray}
&V_z^T=\l z,~ zc_{\te},~ zs_{\te},~ z,~ zc_{\varphi },~ zs_{\varphi }e^{-i\om }\r , \nonumber \\
&\lan {\cal K}(z)\ran^T =\fr{1}{1-4z^2}{\bf 1}_{6\tm 6}+\fr{z^2}{(1-4z^2)^2}\tm \nonumber \\
&\!\!\left( \!\!\!
  \begin{array}{cccccc}
    1 & \!c_{\te} & s_{\te} &\! 1 & c_{\varphi } &\! s_{\varphi }e^{-i\om } \\
   c_{\te} &\! c_{\te}^2 & c_{\te}s_{\te} &\! c_{\te} & c_{\te}c_{\varphi } & \!c_{\te}s_{\varphi }e^{-i\om } \\
   s_{\te} & \!c_{\te}s_{\te} & s_{\te}^2 &\! s_{\te} & s_{\te}c_{\varphi } & \!s_{\te}s_{\varphi }e^{-i\om } \\
    1 & \!c_{\te} & s_{\te}& \!1 & c_{\varphi }  & \!s_{\varphi }e^{-i\om }  \\
    c_{\varphi } &\! c_{\te}c_{\varphi } & s_{\te}c_{\varphi } &\! c_{\varphi } & c_{\varphi }^2 & \!c_{\varphi }s_{\varphi }e^{-i\om }  \\
    s_{\varphi }e^{i\om } &\! c_{\te}s_{\varphi }e^{i\om } & s_{\te}s_{\varphi }e^{i\om }  &\!s_{\varphi }e^{i\om }  &
     c_{\varphi }s_{\varphi }e^{i\om }  & \!s_{\varphi }^2 \\
  \end{array}
\!\!\!\right)~~~~
\la{VzK}
\end{eqnarray}
Using (\ref{VzK}) in (\ref{kin-z}) and introducing canonically normalized real scalar  $\phi $ - the inflaton -
we obtain
\begin{eqnarray}
&{\cal K}_{I\bar J}\pl \Phi_I \pl \Phi_J^{*}\to 4\fr{(\pl z)^2}{(1-4z^2)^2}\equiv \fr{1}{2}(\pl \phi)^2 .
\la{kin}
\end{eqnarray}
From the last equality of (\ref{kin}) we can get the following relation
\begin{eqnarray}
&z=\fr{1}{2}\tanh (\fr{\phi }{\sqrt{2}}) ~,
\la{z-t}
\end{eqnarray}
where $\phi $ is canonically normalized inflaton field.
Moreover, due to the form of the ${\cal K}$ in (\ref{totalK}) and the VEV configuration (\ref{VEVs}),
we have $e^{\cal K} {\cal K}^{{e^{-}}^\dag e^{-}}=1$.
 With these, from (\ref{V}), for $\te\ll 1$ (achieved by suitably selecting the value of $\lam $)
  we derive the inflaton potential ${\cal V}$ to have the form:
\begin{eqnarray}
&{\cal V}(\phi )=V_F(\phi )\simeq \fr{\lam_e^2}{16}\tanh^4 (\fr{\phi }{\sqrt{2}}) .
\la{Vinf}
\end{eqnarray}
As we see, tha inflaton potential depends on a single MSSM Yukawa coupling $\lam_e$.
 Its value, i.e. the value of the MSSM parameter $\tan \bt $ \cite{Yetanbt},  will be determined  from $A_s$ - the amplitude of the curvature perturbations.

\section{III. Inflation and Reheating}
\vs{-0.3cm}

The flat shape of the $\tanh \fr{\phi }{\sqrt{2}}$ function for the large values of $\phi $ 
ensures also the flatness of the inflaton potential (\ref{Vinf}).
The dynamics during the slow roll regime is governed by the  slow roll parameters which, derived from the potential -
the  "VSR" parameters - are  \cite{Stewart:1993bc, Liddle:1994dx}:
\beq
\ep =\fr{1}{2}\l \fr{{\cal V}'}{{\cal V}}\r^2,~~
\eta =\fr{{\cal V}''}{{\cal V}},~~
\xi =\fr{{\cal V}'{\cal V}'''}{{\cal V}^2}.
\la{ep-eta}
\eeq
These parameters determine the spectral index  $n_s$, the trnsor-to-scalar ratio $r$
\begin{eqnarray}
&n_s&=1-6\ep_i+2\eta_i +\fr{2}{3}(22-9C)\ep_i^2-
\nonumber \\
& &-(14-4C)\ep_i \eta_i +\fr{2}{3}\eta_i^2+\fr{1}{6}(13-3C)\xi_i ~,
\nonumber \\
&r&=16\ep_i [ 1-(\fr{2}{3}-2C)(2\ep_i-\eta_i) ], ~~ C=0.0815,
\la{r-ns-SRV}
\end{eqnarray}
and the value of the spectral index running
\begin{eqnarray}
 \fr{dn_s}{d\ln k}=16\ep_i\eta_i-24\ep_i^2-2\xi_i ~.
\la{n-run}
\end{eqnarray}
Expressions in Eqs. (\ref{r-ns-SRV}), (\ref{n-run}) are valid within the second order approximation,
which is fully sufficient due to the slow roll regime.
Here and throughout the paper, the  subscript 'i'  indicate that the appropriate
quantity is calculated at point  $\phi_i$, which corresponds to the beginning of inflation.
Similarly,  subscript 'e' will correspond to $\phi_e$ - the point at which inflation ends.

Since the slow roll breaks down at $\phi_e$,  the $\phi_e$'s value should be determined by the exact condition $\ep_H=1$.
The $\ep_H=1$ (the HSR parameter) is derived from the Hubble parameter.
The relations between HSR and VSR parameters (given in Refs. \cite{Stewart:1993bc}, \cite{Liddle:1994dx})
can be used upon analysis of the inflation process.
As far as the $\phi_i$ is concerned, its value (with $\phi_e$ already fixed by the condition $\ep_H=1$) determines  the number
 of $e$-foldings $N_e^{\rm inf}$ during the inflation. The latter is given by the exact expression:
\begin{eqnarray}
&N_e^{\rm inf}=\fr{1}{\sqrt{2}}\int_{\phi_e}^{\phi_i} \fr{1}{\sqrt{\ep_H}} d\phi ~.
\la{exact-Ninf}
\end{eqnarray}
%
%
%
\begin{figure}[!t]
\begin{center}
\includegraphics[width=0.8\columnwidth]{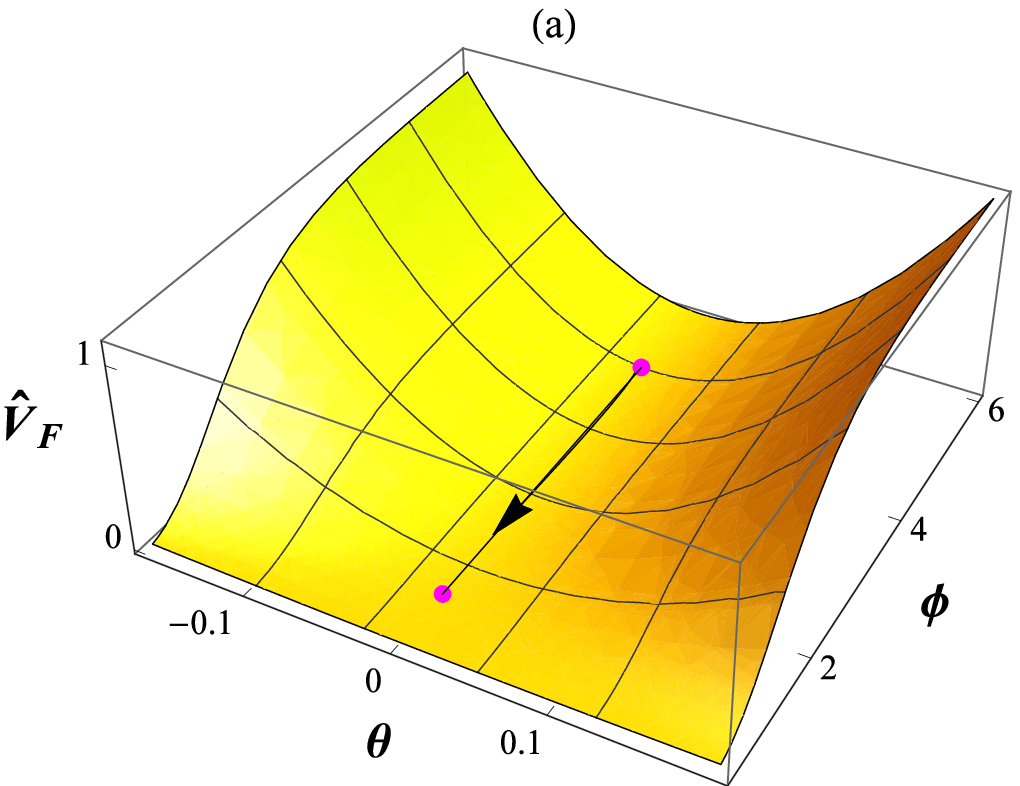}
\vs{0.5cm}
\includegraphics[width=0.8\columnwidth]{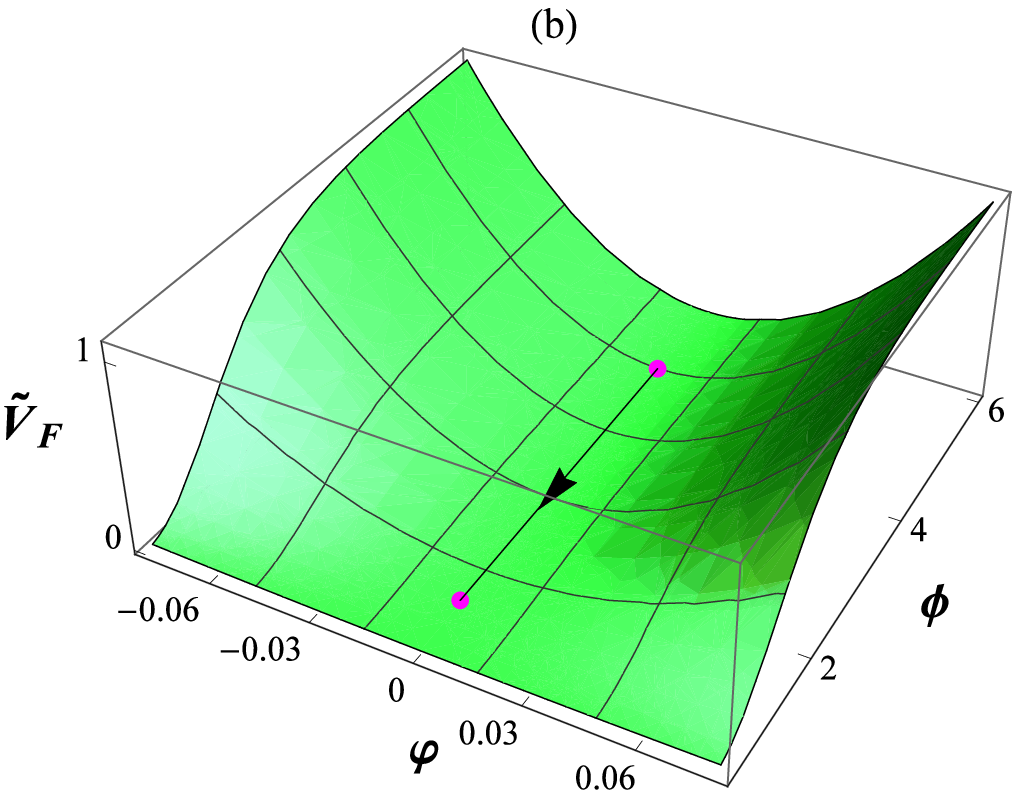}
\caption{{\bf (a):} Dependance of the potential on $\te $ and $\phi $. $\hat{V}_F=V_F/(85\lam_e^2)$
and $\varphi \simeq 3\cdot 10^{-4}$ is taken.
{\bf (b):} Potential as a function of $\varphi $ and $\phi $. $\tilde{V}_F=V_F/(8\lam_e^2)$ and
 $\te \simeq 0.012$ is taken.
Both plots corresponds to the case {\bf (i)} and $\om =\pi+{\rm Arg }\l \fr{V_{ud}}{V_{td}}\r$ is taken.
Arrows correspond to the inflaton's path.
}
\label{plot-1}
\end{center}
\end{figure}
On the other hand, to guarantee the causality of fluctuations,  the $N_e^{\rm inf}$ should satisfy \cite{Liddle:2003as}:
\begin{eqnarray}
&N_e^{\rm inf}=&62-\ln \fr{k}{a_0H_0}-\ln \fr{10^{16}{\rm GeV}}{{\cal V}_i^{1/4}}+\ln \fr{{\cal V}_i^{1/4}}{{\cal V}_e^{1/4}}
 \nonumber \\
&&-\fr{4-3\ga }{3\ga }\ln \fr{{\cal V}_{e}^{1/4}}{\rho_{reh}^{1/4}}~,
\la{Ne}
\end{eqnarray}
where $k=0.002\hspace{0.6mm}{\rm Mpc}^{-1}$  and the present horizon scale $a_0H_0$ is $a_0H_0\approx 0.00033\hspace{0.6mm}{\rm Mpc}^{-1}$.
The factor
$\ga =2\fr{\int_0^{\phi_e} (1-{\cal V}/{\cal V}_e)^{1/2}d\phi }{\int_0^{\phi_e} (1-{\cal V}/{\cal V}_e)^{-1/2}d\phi }$
(equals to$\simeq 1.19$ in our case)
 accounts for the effect of inflaton's oscillation around  its minima  after inflation \cite{Turner:1983he}.
For consistency, we need  to match the values of $N_e^{\rm inf}$ obtained from Eqs. (\ref{exact-Ninf}) and (\ref{Ne}).
As it turns out, within the considered scenario
$\phi_i\simeq 1.1295$ and $\phi_e\simeq 4.8325$ (given in the units of $M_{Pl}=2.4\tm 10^{18}$~GeV).
These points, together with the inflaton's trajectory during the course of the inflations, are shown
in plots of Fig.\ref{plot-1}.
With $\phi_i$ get fixed, we can calculate the observables given in (\ref{r-ns-SRV}) and (\ref{n-run}).
These quantities are calculated by the parameters in (\ref{ep-eta}). The latter are
independent of  the $\lam_e$ -the single coupling appearing in (\ref{Vinf}).  $\lam_e$'s value is important for
the value of the vacuum energy dominantly stored in the scalar potential ${\cal V}$ during the inflation.
 The values  ${\cal V}_{i, e}$ are needed to carry calculations with Eq.  (\ref{Ne}).
On the other hand, another observable - the amplitude of curvature perturbation $A_s$ given by
\begin{eqnarray}
A_s^{1/2}=\fr{1}{\sqrt{12}\pi }\left | \fr{{\cal V}^{3/2}}{M_{Pl}^3{\cal V}'}\right |_{\phi_i}~,
\la{dT-Ts}
\end{eqnarray}
can be used to determine ${\cal V}_{i}$ and consequently the value of $\lam_e$.
In order to get experimentally measured value  $A_s^{1/2}= 4.581 \tm 10^{-5}$ \cite{Akrami:2018odb},
using (\ref{dT-Ts}), we need to have $\lam_e(M_{Pl})= 2.435\tm 10^{-5}$ \cite{phi-val}.
This, in turn allows to determine the MSSM parameter $\tan \bt $ to be \cite{Yetanbt}:
\begin{eqnarray}
\tan \bt \simeq 13.12~.
\la{pred-tan-bt}
\end{eqnarray}

In addition,  calculation of the thermal energy density  $\rho_{\rm reh}\!\!=\!\fr{\pi^2}{30}g_*T_r^4$
is required. It depends on the reheating process (via reheating temperature $T_r$) which is realized by the inflaton's
decay. In this case \cite{Kofman:1997yn}:
\begin{eqnarray}
T_r=\l \fr{90}{\pi^2g_*}\r^{1/4}\!\!\sqrt{M_{Pl}\Ga(\phi )},
\la{r-reh1}
\end{eqnarray}
where $g_*$ is the effective  number of massless degrees of freedon at temperature
$T_r$ ($g_*\!=\!N_b\!+\!\fr{7}{8}N_b$ and in our case is $g_*\!\!=\!42.75$), and $\Ga(\phi )$ is inflaton's decay width.
It turns out that within our model, all parameters involved in the inflation and in this process are known.
This enables us  to calculate $\Ga(\phi )$ and therefore predict the $T_r$.

Since the inflaton comes from the MSSM states, its couplings to the remaining states are well fixed.
The VEV configuration (\ref{VEVs}) breaks the $SU(3)_c\tm SU(2)_L\tm U(1)_Y$ symmetry down to the $SU(2)_c$. Thus, from
the gauge sector only $SU(2)_c$'s states are massless.
Via the Yukawa superpotential, the inflaton field couples to the MSSM chiral superfield states via the $z$ VEV.
 And the very same couplings generate masses (which scale as $z$ times corresponding Yukawa coupling) of the latter states. Because of this, it turns out that states which have tree level
 coupling with the inflaton are heavier than the inflaton. Therefore, inflaton's tree level decays are either kinematically forbidden
 or (if realized via many body decays) are strongly suppressed.

The dominant decay of the inflaton  $\phi $ happens radiatively (via 1-loop correction) in two massless gluons of
the unbroken $SU(2)_c$.
Corresponding decay width is given by:
\begin{eqnarray}
\label{inf-width}
&\Ga (\phi )\!\simeq \!\Ga (\phi \to gg)\!= \!
\fr{3m_{\phi }^3\al_s^2}{2(8 \pi )^3}\left |\sum_Qf_QA_{1/2}(\tau_Q) \right |^2
 \nonumber \\
& \simeq \fr{m_{\phi }^3\al_s^2}{48\pi^3}\l \fr{F'}{F}+\fr{F_g'}{F_g}\r^2~,
\end{eqnarray}
where
$m_{\phi }^2={\cal V}''$ (is the inflaton's mass), $\tau_q=\fr{m_{\phi }^2}{4m_Q^2}$ and
$m_Q$ denote masses of $SU(2)_c$ colored fermions which couple with the inflaton.
 Among them are $SU(2)_c$ doublets from massive $s, b$ quarks, which circulate into the loop diagram.
 For them $f_{s,b}\!=\!\fr{F'}{F}$ is taken in Eq. (\ref{inf-width}) .
 Their canonically normalized couplings to the inflaton emerges from the Yukawa term:
\begin{eqnarray}
&\fr{1}{2}F(\phi )d^TY_Dd^c,~&
F(\phi )\!=\!\tanh \!\fr{\phi }{\sqrt{2}} (1\!-\!\tanh^2 \!\!\fr{\phi }{\sqrt{2}})^{1/2},~
\la{inf-ff}
\end{eqnarray}
which for $\phi $-$d$-$d^c$-type interaction gives
$\fr{1}{2}F'\phi d^TY_Dd^c$, where $F'=\fr{dF}{d \phi}$, and that's how the term $\fr{F'}{F}=f_{s,b}$ appears
 in Eq. (\ref{inf-width}).
 Besides these, in the loop (governing inflaton's decay) two
 massive Dirac fermions circulate, which are formed after $SU(3)_c\to SU(2)_c$ breaking and pairing
corresponding gauginos and colored matter. For them
$f_{g}\!=\!\fr{F_g'}{F_g}$ was  used in (\ref{inf-width}) with $F_g(\phi )\!=\!\sinh \!\fr{\phi }{\sqrt{2}}$.
  The function $A_{1/2}(\tau_Q)$ in Eq. (\ref{inf-width})
has property $\left.A_{1/2}(\tau_Q)\right |_{\tau_Q\ll 1}\simeq 4/3$ \cite{Djouadi:2005gi}.
In (\ref{inf-width}) all $\phi $ dependent quantities need to be evaluated at point $\phi =\phi_e$.

Having all these expressions, we can carry out detailed analysis related to the inflation process. Doing so,
for the observables we obtain \cite{phi-val}:
\begin{eqnarray}
\label{result-sum}
&n_s=0.9662 , ~~r=0.00118 ,~~\fr{dn_s}{d\ln k}=-5.98\cdot 10^{-4},
 \nonumber \\
&N_e^{\rm inf}=57.74 , ~~~\rho_{\rm reh}^{1/4}=2.61\cdot 10^{7}{\rm GeV},
\nonumber \\
&T_r=1.35\cdot 10^{7}{\rm GeV} .
\end{eqnarray}
As one can see, the values of $n_s, r$ and $\fr{dn_s}{d\ln k}$ are in good agreement
with the current observations \cite{Akrami:2018odb}. We will comment
about the value of the reheating temperature $T_r$  in the next section, where some implications and
related phenomenology are discussed.

\section{IV. Related Phenomenology and Discussion}
\vs{-0.3cm}

In this section, first we discuss some implications and phenomenology related to the inflationary scenario we have
presented above and then give brief summary.

\vs{-0.3cm}
\subsection{IV.1.  Relic gravitinos}
\vs{-0.3cm}
For the reheat temperature  $T_r$ obtained in this schenario [see Eq. (\ref{result-sum})],
the thermally produced gravitino abundance can  easily be compatible \cite{TR-bounds} with observations
for specific and phenomenologically viable sparticle spectroscopy.

As far as the non-thermal gravitino production, via the inflaton decay is concerned, as shown in \cite{Nilles:2001fg},
this process can be adequately suppressed. However, results of \cite{Nilles:2001fg} applies for minimal K\" ahler potential.
If nonminimal K\" ahler potential involves specific mixing terms between the inflaton $z$ (as denoted in the present work) and SUSY breaking superfield $X$,
then situation in general would be different \cite{Dine:2006ii,Endo:2006tf,Kawasaki:2006hm}. As was pointed out \cite{Dine:2006ii,Kawasaki:2006hm},
 the additional $\de {\cal K}=|z|^2X^2$  K\" ahler potential coupling,
can lead to the gravitino overproduction. This term can be easily forbidden if $X$ field transforms either under some $U(1)$
or $R$-symmetry, or under discrete symmetry (such as for instance $Z_3$).
Thus, the details of the SUSY breaking sector is important.
On the other hand,  connection of the SUSY breaking mechanism with our inflation model
 deserves separate investigation.

\vs{-0.3cm}
\subsection{IV.2.   Neutrino masses}
\vs{-0.3cm}
Within the considered scenario, in case {\bf (i)} [see paragraph after Eq. (\ref{Fterms})] the
lepton number violating $W'=-\lam q_1l_2d^c$
superpotential coupling, which also breaks matter parity, was exploited. This, at 1-loop level induce $\mu_ih_ul_i$-type superpotential
and soft $B_ih_u\tl l_i$ terms, which result neutrino mass
$m_{\nu_{\mu}}\!\approx \!\fr{\lam^2g_2^2}{4c_w^2}\fr{m_d^2}{\tl m}\!\l \!\fr{9}{8\pi^2}\ln \fr{M_{Pl}}{M_Z}\!\r^2$ \cite{Rpar}, by neutralino exchange [similar to seesaw induced neutrino mass, generated by the right handed neutrino (RHN) exchange], where $\tl m$ is the SUSY scale (for simplicity we have assumed that all sparticles have masses close to $\tl m$).
 This, by demanding
$m_{\nu }\stackrel{<}{\sim }0.1$~eV for $\tl m=2$~TeV gives the bound $\lam \stackrel{<}{\sim }0.1$. This, together with
desirably suppressed value of $\tan \te \simeq \fr{\lam_d}{\lam }<0.1$ gives
$6\tm 10^{-4}\stackrel{<}{\sim }\lam \stackrel{<}{\sim }0.1 $.
The $\lam q_1l_2d^c$ superpotential coupling term also directly contribute to the 1-loop neutrino
 mass$\sim \fr{3\lam^2}{8\pi^2}\fr{m_d^2}{\tl m}$ \cite{Rpar}. This, for $\lam \stackrel{<}{\sim }0.1 $ gives
 more suppressed contribution $\de m_{\nu }\stackrel{<}{\sim }2\tm 10^{-3}$~eV.
 Although these neutrino mass scales are close to the values needed for accommodation of the neutrino data \cite{nu-data},
 by the $W'=-\lam q_1l_2d^c$ coupling alone would be hard and challenging to get also desirable neutrino mixing pattern.
  To achieve all these, one way is to include additional $\bar \lam_{ijk}e^c_il_jl_k$ and $\lam_{ijk}q_il_jd^c_k$-type terms, and by proper selection of various couplings obtain consistent neutrino sector.
   However, within our study, one should also take care that considered inflation model remains intact.
Alternative, and perhaps simpler, way would be to include  RHN state(s), which can lead to desirable neutrino oscillations via the
 contribution of conventional seesaw mechanism \cite{see-saw}. This possibility definitely seems the simplest
choice especially for our case {\bf (ii)}, which preserves matter parity and lepton number. Detailed study of the neutrino sector
in connection to the considered model of inflation should be pursued elsewhere.

\vs{0.3cm}
Concluding, within the MSSM we have presented model of inflation in which the inflaton is a combination of the Higgs, slepton
and squark states. While the VEVs of these states are along the flat $D$-term trajectory, the inflation is driven
by the vacuum energy of the electron Yukawa superpotential. This uniquely fixes the value of the MSSM parameter $\tan \bt $
[see Eq. (\ref{pred-tan-bt})]. To our knowledge, it is first example with such close connection between
the particle physics model and inflationary cosmology.
 Since all parameters involved in the inflation and postinflationary reheating processes
 were known, the presented model is very predictive.

Encouraged by these findings, would be interesting to realise similar constructions in a framework of other well
motivated SUSY constructions such as left-right symmetric and grand unified  [i.e. $SU(5)$, $SO(10)$, etc.] models.
Note that, if within the GUTs, inflaton condensate (being either Higgs, slepton or squark state) breaks the GUT symmetry,
then (as shown in Ref. \cite{Dvali:1998ct}) within such construction the monopole problem can be easily avoided.
Investigation of these  exciting issues  will be performed elsewhere.

\begin{acknowledgments}
\end{acknowledgments}
%

%


\begin{thebibliography}{26}
\bibitem{phi-val}
Since within the considered scenario the inflation occurs at high scales (with large values of the inflaton),
upon the calculations we use the RG evaluated values of the Yukawa couplings, masses and the CKM matrix elements at scale $M_{Pl}$.
\bibitem{te-approx}
Strictly speaking, the value of $\te $ should be found from the minimization of the whole potential.
However, since for both cases we consider $\te \ll 1$, with improved accuracy we will have
$\tan \te\!\simeq \!\fr{\lam_d}{\lam }\l 1+\fr{\lam_d^2+\lam_e^2}{\lam^2}\r$ and
$\sin \te \!\simeq \!\fr{\lam_d}{\lam z^4}\l 1+\fr{\lam_d^2+\lam_e^2}{(\lam z^4)^2}\r$ for cases
 {\bf (i)} and  {\bf (ii)} respectively. For $\fr{\lam_d}{\lam }<0.1$ and $\fr{\lam_d}{\lam z^4}<0.1$ additional corrections
 are$\stackrel{<}{_\sim }{\rm 1}\%$ and we do not need to include them in our considerations because
 in practice they are insignificunt.
\bibitem{Yetanbt}
Near the SUSY scale, between the $\lam_e$ and $\tan \bt $ we have the relation:
$\lam_e=\fr{m_e}{v}\sqrt{1+\tan^{-2}\bt }$.
\end{thebibliography}
\end{document}